\documentclass[onecolumn,british,english,pra,superscriptaddress,groupedaddress]{revtex4}

\usepackage{amsfonts}
\usepackage{amsmath}
\usepackage{graphicx}
\usepackage[utf8]{inputenc}
\usepackage{amssymb}
\usepackage{times}
\usepackage{color}

\begin{document}

\title{Relative multiplexing for minimizing switching in linear-optical quantum computing}

\author{Mercedes Gimeno-Segovia}
\affiliation{Quantum Engineering Technology Labs, H. H. Wills Physics Laboratory and Department of Electrical and Electronic Engineering, University of Bristol, BS8 1FD, UK}
\affiliation{Institute for Quantum Science and Technology, University of Calgary, Alberta T2N 1N4, Canada}

\author{Hugo Cable}
\affiliation{Quantum Engineering Technology Labs, H. H. Wills Physics Laboratory and Department of Electrical and Electronic Engineering, University of Bristol, BS8 1FD, UK}

\author{Gabriel J. Mendoza}
\thanks{Current address: Hewlett Packard Labs, 1501 Page Mill Rd., Palo Alto, CA, US}
\affiliation{Quantum Engineering Technology Labs, H. H. Wills Physics Laboratory and Department of Electrical and Electronic Engineering, University of Bristol, BS8 1FD, UK}

\author{Pete Shadbolt}
\affiliation{Department of Physics, Imperial College London, London SW7 2AZ, United Kingdom}

\author{Joshua W. Silverstone}
\affiliation{Quantum Engineering Technology Labs, H. H. Wills Physics Laboratory and Department of Electrical and Electronic Engineering, University of Bristol, BS8 1FD, UK}

\author{Jacques Carolan}
\affiliation{Department of Electrical Engineering and Computer Science, Massachusetts Institute of Technology, Cambridge, Massachusetts 02139, United States}

\author{Mark G. Thompson}
\affiliation{Quantum Engineering Technology Labs, H. H. Wills Physics Laboratory and Department of Electrical and Electronic Engineering, University of Bristol, BS8 1FD, UK}

\author{Jeremy L. O'Brien}
\affiliation{Quantum Engineering Technology Labs, H. H. Wills Physics Laboratory and Department of Electrical and Electronic Engineering, University of Bristol, BS8 1FD, UK}

\author{Terry Rudolph}
\affiliation{Department of Physics, Imperial College London, London SW7 2AZ, United Kingdom}

\begin{abstract}
Many existing schemes for linear-optical quantum computing (LOQC) depend on \emph{multiplexing} (MUX), which uses dynamic routing to enable near-deterministic gates and sources to be constructed using heralded, probabilistic primitives. MUXing accounts for the overwhelming majority of active switching demands in current LOQC architectures. In this manuscript we introduce \emph{relative multiplexing} (RMUX), a general-purpose optimisation which can dramatically reduce the active switching requirements for MUX in LOQC, and thereby reduce hardware complexity and energy consumption, as well as relaxing demands on performance for various photonic components.  We discuss the application of RMUX to the generation of entangled states from probabilistic single-photon sources, and argue that an order of magnitude improvement in the rate of generation of Bell states can be achieved.  In addition, we apply RMUX to the proposal for percolation of a 3D cluster state in [PRL 115, 020502 (2015)], and we find that RMUX allows an 2.4x increase in loss tolerance for this architecture.
\end{abstract}

\maketitle

\section{Introduction}
\label{sec:Introduction}

A compelling approach to quantum information processing is provided by linear-optical quantum computing (LOQC), where information is encoded in photonic qubits and gates are implemented using the standard toolkit of linear-optic experiments.  In 2001, Knill, Laflamme and Milburn proved theoretically that single-photon sources (SPS's), passive linear-optical components, photon-number-counting detectors and, feedforward measurements incorporating active switching, are sufficient in principle to enable universal quantum computing \cite{Knill2001}.  Since then, several proposals have substantially improved upon their approach \cite{Kok2007, Rudolph2016}.  By exploiting the paradigm of measurement-based quantum computing (MBQC) \cite{Raussendorf2001}, it has been shown that orders of magnitude reductions in resource counts are possible compared to \cite{Knill2001} for LOQC based on single-photon encodings of qubits \cite{Nielsen2004, Browne2005, Gimeno-Segovia2015}.

In parallel with these theoretical developments, the emergence of the field of integrated quantum photonics has led to the demonstration of reconfigurable waveguide circuits which achieve high-visibility quantum interference in multi-photon experiments \cite{Shadbolt2012,Carolan2015}.  Recent demonstrations \cite{Silverstone2015,Harris2014} show tremendous potential for integrated devices using a platform such as silicon photonics which can support high component densities, and which may eventually enable large-scale implementation of LOQC \cite{Rudolph2016}.  However, stochasticity is intrinsic to all architectures for LOQC which work at the single-photon level; it creates major challenges for experimental implementation and scalability, as the integration of active switching which is simultaneously ultra-fast, low loss and low noise has yet to be achieved in any photonic hardware platform.

There are two main sources of stochasticity in LOQC: Firstly, there are currently no on-demand deterministic sources (in particular of single photons \cite{Eisaman2011} or Bell pairs \cite{Zhang2008,Barz2010}) that meet all requirements for large-scale LOQC, namely stringent requirements for photon indistinguishability, high purity, low noise, and ready compatibility with integrated photonic circuitry.  High-purity photons can be generated using sources based on spontaneous parametric downconversion or spontaneous four-wave mixing, but these methods are fundamentally probabilistic \cite{Gerry2005}.  Secondly, all linear-optical entangling operations for the standard dual rail (spatial or polarisation) qubit encodings are fundamentally non-deterministic \cite{Kok2007}, which affects schemes for generating entanglement \cite{Zhang2008,Varnava2008} and for performing (incomplete) Bell measurements \cite{Lutkenhaus1999,Grice2011,Ewert2014}.

One way to achieve scalability while using stochastic sources and circuits is to employ multiplexing (MUX) i.e. to repeat non-deterministic operations in parallel (either spatially or temporally) and to integrate all outcomes via a switching network as successful events are ``heralded''.  A substantial body of theoretical \cite{Migdall2002,Jeffrey2004,Shapiro2007,McCusker2009,Mower2011,Broome2011,Jennewein2011,Christ2012,Glebov2013,Mazzarella2013,Schmiegelow2014,Bonneau2015} and experimental \cite{Ma2011,Collins2013,Meany2014,Mendoza2015,Kaneda2015,Xiong2016} research focuses on using multiplexing to improve single-photon generation performance.  Furthermore, complicated multiplexing schemes could in principle enable the implementation of LOQC based upon repeat-until-success strategies \cite{Varnava2008,Li2015}.  An alternative to repeat-until-success methods is a ballistic approach for which active switching is not required for the process of cluster state generation \cite{Kieling2007}.  However the current leading proposal along these lines requires (near)-deterministic three-photon GHZ (3-GHZ) states at the start \cite{Gimeno-Segovia2015}, and multiplexing techniques would be needed to generate these resource states.

Since active switching will most likely represent a dominant source of losses (and other forms of decoherence) in future experiments, we explore a general technique that we term {\it relative multiplexing} (RMUX), with the aim of minimizing requirements for active switching used in multiplexing throughout LOQC architectures.  In section \ref{sec:BasicMultiplexing}, we discuss some standard multiplexing methods and argue how they become inefficient when used in concatenated schemes.  We introduce the key idea of RMUX in section \ref{sec:RMUX}, and analyze how best to synchronise events when using this new type of multiplexing.  In section \ref{sec:BallisticArchitecture}, we discuss one way in which a RMUX strategy can be applied to the ballistic architecture of \cite{Gimeno-Segovia2015}, for which a 3d cluster state is generated on a diamond lattice using 3-GHZ states, and entangling gates which operate with success probability above the corresponding percolation threshold. This leads to improved tolerance to photon loss which we explain in section \ref{sec:PercolationResults}, before concluding in section \ref{sec:Conclusions}.

\section{Spatial and temporal multiplexing}
\label{sec:BasicMultiplexing}

\begin{figure}[t]
\begin{center}
\includegraphics[width=0.9\linewidth]{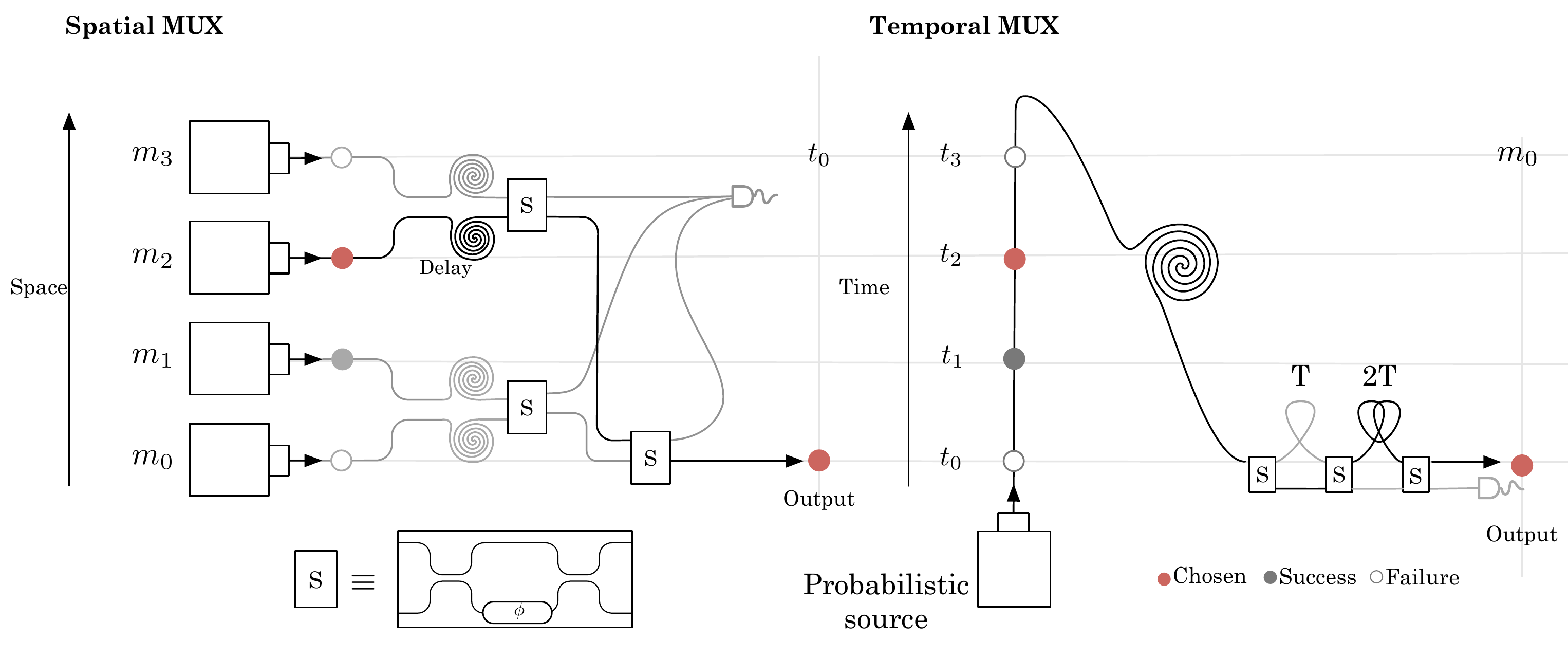}
\caption{{\it Schematic layout for spatial and temporal multiplexing ($k=4$):} Note that the axes for space and time have been interchanged in the figures in order to highlight the equivalence of both methods. The sources emit one photon in the spatiotemporal mode $m_0 t_0$. In spatial MUX a $k\times 1$ switch (realised here with cascaded Mach-Zehnder interferometers) locates the emitted photon in mode $m_0$, while in temporal MUX a reconfigurable delay line is used to change the temporal mode $t_0$. \textcolor{black}{The maximum delay that can be achieved with either switching network is $2^{s-1}-1$, where s is the number of $2\times 2$ switches used.} }
\label{fig:SpatialTemporalMUX}
\end{center}
\end{figure}

Arrays of non-deterministic heralded single-photon sources (HSPS), each with efficiency (per pulse) of $\eta$, are not suitable for LOQC on their own since the probability of simultaneously generating $n$ indistinguishable photons decreases exponentially as $\eta^n$.  In figure~\ref{fig:SpatialTemporalMUX}, we illustrate examples of spatial and temporal MUX sources which circumvent this problem by using repeated source generation and fast active reconfiguration, to relocate photons in desired ``logical'' spatio-temporal bins.  In the spatial MUX scheme, $k$ HSPS's are pumped simultaneously, emitting into a single time bin but different spatial modes. Upon heralding of success, the switch configuration is set to redirect one photon to the output mode while the photons are stored in delay lines. The selected photon is then directed to the output port while the extraneous photons are re-routed to a detector or beam dump.  In the temporal MUX scheme, a probabilistic HSPS is pumped $k$ times to generate a series of events in different time bins and the same spatial mode. The switch network is configured to select a particular delay of between $0$ and $k-1$ (in addition to passive delay for switch reconfiguration) to locate one of the photons in a particular temporal bin, while the extraneous photons are again discarded.

When device imperfections can be ignored, these multiplexing schemes can boost the success probability from $\eta$ to $p_s$ or better using a numbers of repetitions $k$ satisfying,
 \begin{equation}
1-(1-\eta)^k\geq p_s,
\label{eq:mux}
\end{equation}
and in principle $p_s$ can approach unity.  An analysis which includes the effects of component losses and detector inefficiencies is given in \cite{Bonneau2015} for various alternative switching architectures (using threshold or number-resolving photon detectors).

{\it Log-tree and binary-delay schemes:} The multiplexing methods we discuss in this paper require reconfigurable switch arrays to redirect from one of $k$ input bins to one specific spatio-temporal bin.  We will primarily consider $k \times 1$ switch networks constructed using a logarithmic tree (spatial MUX) or binary delay (temporal MUX) networks of $2 \times 2$ switches, which are illustrated for $k=4$ in figure~\ref{fig:SpatialTemporalMUX}. The $2 \times 2$ switches can be implemented using a Mach-Zehnder interferometer with a controllable phase-shifter.  Defining $k^\uparrow =2^{\lceil \log_2(k)\rceil}$, which corresponds to a rounding of $k$ to a (larger) value achievable using a log-tree or binary-delay scheme, we can see that these switch networks have depth $1+\log_2(k^\uparrow)$. \textcolor{black}{The maximum delay that can be achieved with either switching network is $2^{s-1}-1$, where s is the number of $2\times 2$ switches used.} In practical terms, spatial MUX has the disadvantage of needing a large number of redundant sources, and hence additional circuitry on a photonic chip.  Temporal MUX however has the disadvantage of needing longer delays, including a requirement for multiple delays lines of various sizes.  It also reduces the effective clock rate by a factor of $k$.  The need for fast reconfigurability with both types of MUX imposes severe technological restrictions on the switches.  However from a theoretical point of view, temporal and spatial MUX are conceptually equivalent.  In the rest of the paper, we will restrict our discussion to temporal schemes (typically with binary-delay networks) but with the understanding that analogous statements apply for spatial MUX.

{\it Inefficiency in concatenated multiplexing schemes:}  Schemes for LOQC typically demand multiple stages of multiplexing, for example to generate deterministic single photons, entangled resource states from single photons, and finally large quantum states from the resource states \cite{Gimeno-Segovia2015,Li2015}.  When designing these more complicated multiplexing schemes, the goal is typically to achieve a success probability $p_s$ close to $1$, and the numbers of interest are the repetitions and the size of the switching network needed to achieve this.  Generally however, a large average number of successful events must be discarded to achieve high values for $p_s$.

For example, let us consider the generation of a single 3-GHZ state from an array of HSPS's using the scheme of \cite{Varnava2008}, which (non-deterministically) generates a 3-GHZ state from six single photons at the input.  We will assume two stages of MUX: at the first stage, six MUX sources are required to increase the single-photon emission rate from $\eta=0.1$ (which is typical for sources using spontaneous parametric downversion) to $p_1=0.99$; at the second stage, the GHZ generator itself is multiplexed to boost its success rate from $1/32$ (which assumes all six photons are delivered at the input) to $p_2=0.99$.  Details of this MUX scheme are shown in table~\ref{table:GHZMUX} assuming a binary-delay network at both levels, and the potential generation of quantum states at each stage is also shown.  Overall the scheme generates one GHZ state with $p_s=0.93$.  However overall enough single photons are generated to attempt generation of a 3-GHZ 1638.4 times, on average generating 51.2 GHZ states.
\begin{table}
\centering
\setlength\tabcolsep{10.5pt}
\begin{tabular}{c c c c c c}
\hline \hline
 &Initial prob.& Post MUX prob. &Bins $k(k^\uparrow) $& Switch depth & Potential resources (mean)\\
\hline \hline
Stage 1: HSPS       & 0.1          & $p_1=0.99$     & 44(64)  & 7 & 6.4 photons   \\ \hline
Stage 2:  & 1/32=0.03125 & $p_2=0.99$     & 146(256)& 9 & 8 GHZ   \\
3-GHZ from 6 photons  &              &                &         &   &               \\ \hline
Stages 1 and 2 combined: &              & $p_s=0.99^7$   & (16384) & 16& 9830.4 photons\\
3-GHZ from 6 HSPSs   &              & $\,\,\,\,=0.93$&         &   & 51.2 GHZ      \\
\hline
\end{tabular}
\caption{{\it Resources to generate one 3-GHZ state.}
The process of generating a 3-GHZ state with high probability is broken down into steps: the first generating a near-deterministic source of single photons from HSPSs, the second generating a near-deterministic source of 3-GHZ states assuming deterministic single photons at input and the third generating a near-deterministic 3-GHZ state from HSPSs, which is a combination of the two first steps.}
\label{table:GHZMUX}
\end{table}

\begin{figure}[hbt]
\begin{center}
\includegraphics[width=\linewidth]{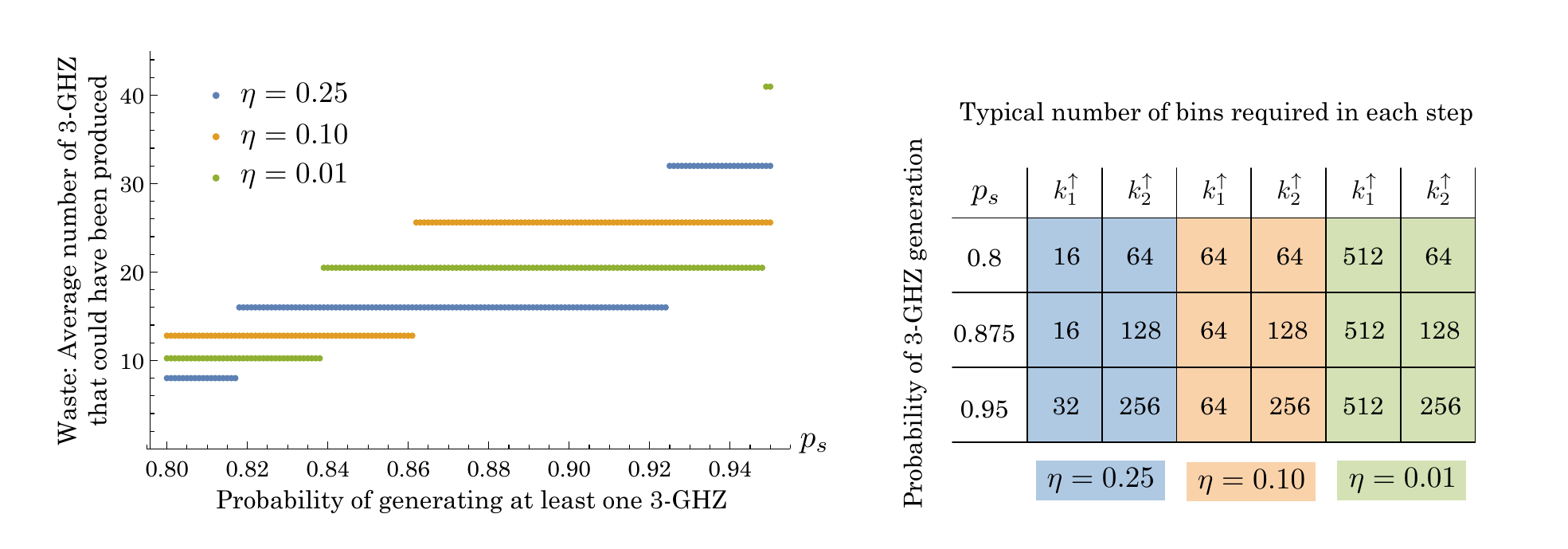}
\caption{
{\it Unused potential for 3-GHZ generation using concatenated MUX:}
Number of 3-GHZ states that could have been produced on average when using a MUX scheme with probability $p_s$ for generating one near-deterministic 3-GHZ state. These results are calculated for three different source efficiencies, indicated by the three colours of the graph. The table on the right indicates typical values of $k_1^{\uparrow}$ and $k_2^{\uparrow}$ necessary to achieve the probability of generating a single 3-GHZ state, $p_s$.}
\label{fig:aGHZ}
\end{center}
\end{figure}

The number of surplus states that are wasted depends on the source efficiency and the required overall probability of success for the 3-GHZ states. We can repeat the calculation above to obtain the number of extra resource states that could have been produced on average when attempting to produce a single 3-GHZ for various values of $p_s$. Figure \ref{fig:aGHZ} shows the results of this comparison, where we present the most economical strategy by minimising over $p_1$ and $p_2$: the strategy is optimised to waste the least amount of resource states per \emph{``deterministic"} 3-GHZ. Each multiplexing stage is required to have an output probability of $p_i \geq 0.8$.  It is interesting to note that for the same probability of generating at least one 3-GHZ state, in some cases a source with $1\%$  efficiency would generate a higher number of unused resources than a source with $25\%$ efficiency but less than the source with $10\%$ efficiency (for example in the case $p_s=0.90$). This is counterintuitive because we would expect a trend; however, the average number of 3-GHZ that could have been produced is the multiplication of the source efficiency (to the power of the number of photons used, i.e. six) times the number of bins due to the multiplexing. For a source with very low efficiency, the number of bins increases dramatically. For example, for the case detailed in table \ref{table:GHZMUX} ($p_s=0.93$), we have that a $\eta=0.1$ efficiency source would have a total bin count (time bins across all streams) of $9.8\cdot 10^4$, an $\eta=0.01$ source would have $7.9\cdot 10^5$ and the $\eta=0.001$ source would have $1.3\cdot 10^7$ bins, which shows an increase of two orders of magnitude of the number of bins with respect to the $\eta=0.01$ source, but their efficiency differs only by one order of magnitude. It is also worth noting that the typical values of $k_2^{\uparrow}$ are independent of source efficiency, while values for $k_1^{\uparrow}$ are strongly dependent on source efficiency, as expected.

Physical constraints dictate that not all spare photons generated in the course of multiplexing can be used effectively, as this would require the ability to synchronise any subset of photons at will. In the remainder of this manuscript, we develop the RMUX approach to achieve better (and in some cases optimal) strategies for utilising non-deterministic resources in complicated linear-optic circuits.

\section{Relative multiplexing and synchronising streams of events}
\label{sec:RMUX}

Traditional multiplexing attempts relocation to one fixed spatial mode or temporal bin, and we will refer to this type of multiplexing as ``Standard MUX''. If we were to only require the generation of a successful event in \emph{any} spatio-temporal mode, we would not need to use active switching at all, only knowledge of where the event is located. In an LOQC architecture, the only reason to change the spatial mode or time bin is to synchronise with other events. For example, fusion gates \cite{Browne2005} are based on Hong-Ou-Mandel interference and require photons at the input to be indistinguishable---including for arrival time, and hence require synchronisation. However, changing the spatial mode or time bin of one of two photons undergoing quantum interference, rather than both, is typically sufficient.  This relative synchronisation or co-location of events is the goal of RMUX. \textcolor{black}{One key difference we must point out between these schemes is that, while Standard MUX requires that the maximum delay is determined by the probability of emission of the source (sources with low probability of emission requiring large delays), in RMUX, we can use networks with a much lower number of switches for sources with the same probability of emission.}

\begin{figure}[t]
\includegraphics[width=0.5\columnwidth]{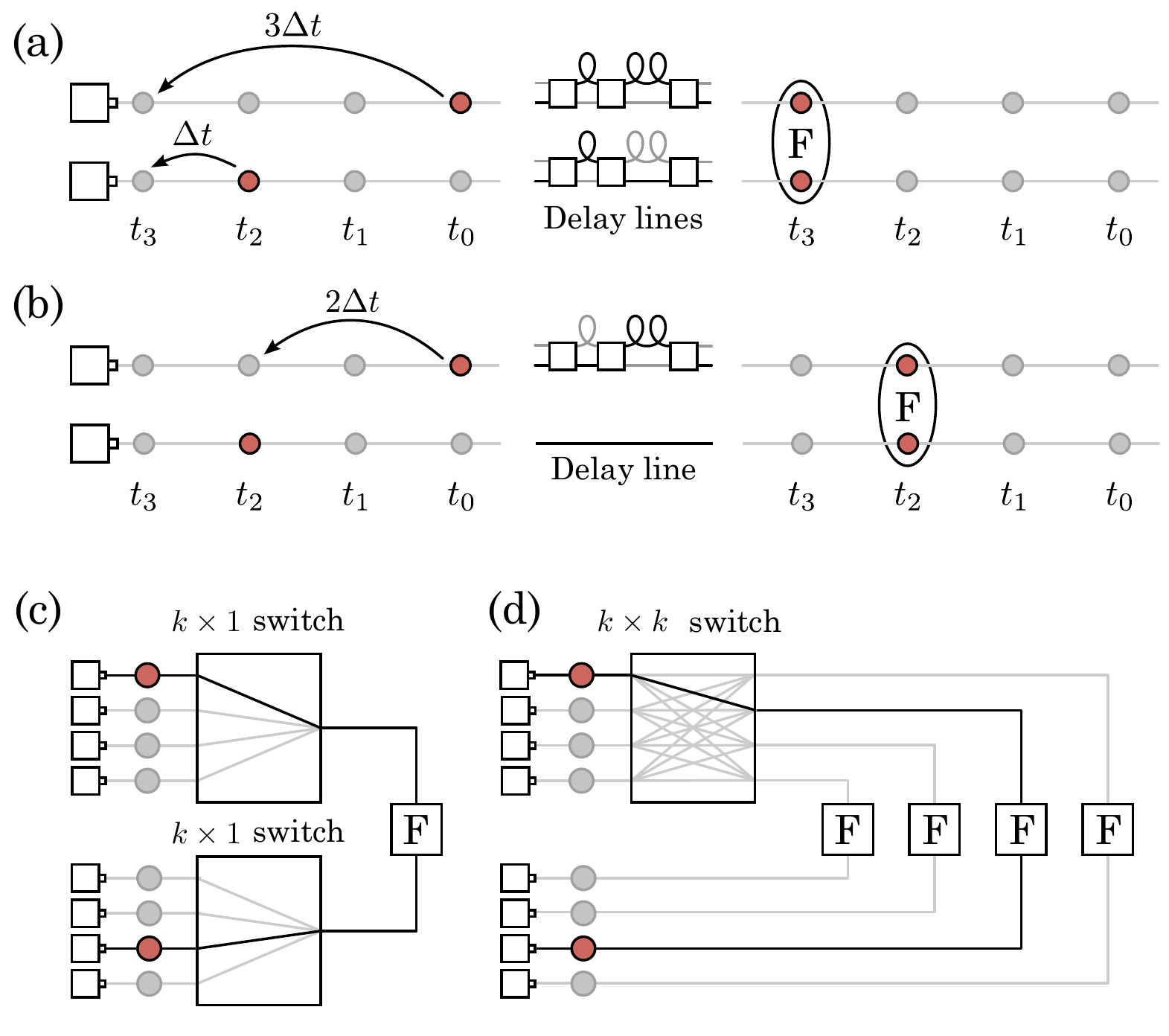}
\caption{{\it Comparison of RMUX with standard MUX before a fusion operation (F):} (a) Temporal standard MUX: two photons are synchronised by delaying both to the same time bin using two binary-delay networks. (b) Temporal RMUX: two photons are synchronised by delaying the photon ahead in time to the time bin of the second photon using one binary-delay network.  Note it does not promote a photon ahead in time. (c) Spatial standard MUX: two photons are rerouted, each through an $k\times 1$ switch, to two specific spatial modes. (d) Spatial RMUX: a $k\times k$ switch is used to relocate one photon in a spatial mode paired with that for the second photon.}
\label{fig:relative_multiplexing}
\end{figure}
Figure \ref{fig:relative_multiplexing} shows the simplest schemes for RMUX as compared to standard MUX, for temporal synchronisation and spatial relocation of pairs of photons ahead of a fusion operation.  As is clear from the figure, the temporal and spatial variants work in close analogy.  Hence, without loss of generality, we will again limit the discussion to temporal RMUX.  Using RMUX, events do not have to be synchronised to an overall clock cycle, but only with respect to a limited number of other events.  While RMUX is conceptually more complex than standard MUX, it allows for a better usage of resources and less stringent requirements on optical components.  This is crucial for implementing LOQC architecture for which there are stringent constraints on component specifications.

{\it Optimal synchronisation of two streams of events using RMUX:}  As the basic building block for RMUX schemes, we can investigate the problem of optimally matching events generated probabilistically in two ``abstract'' streams.  Once solved, this building block can be used to analyse matching on any number of streams, using for example a cascade of RMUX schemes on successive pairs of streams, as we discuss at the end of this section.  We now consider two streams which represent photons generated from two independent sources, where both sources generate a photon with success probability $p$.  Our aim is to find the optimal set of delays to apply to photons in the first stream, so as to synchronise as many photons as possible in the two streams, and with as little overall delay as possible (thus minimising additional losses and possible decoherence from delay lines).

\begin{figure}[t]
\includegraphics[width=0.9\linewidth]
{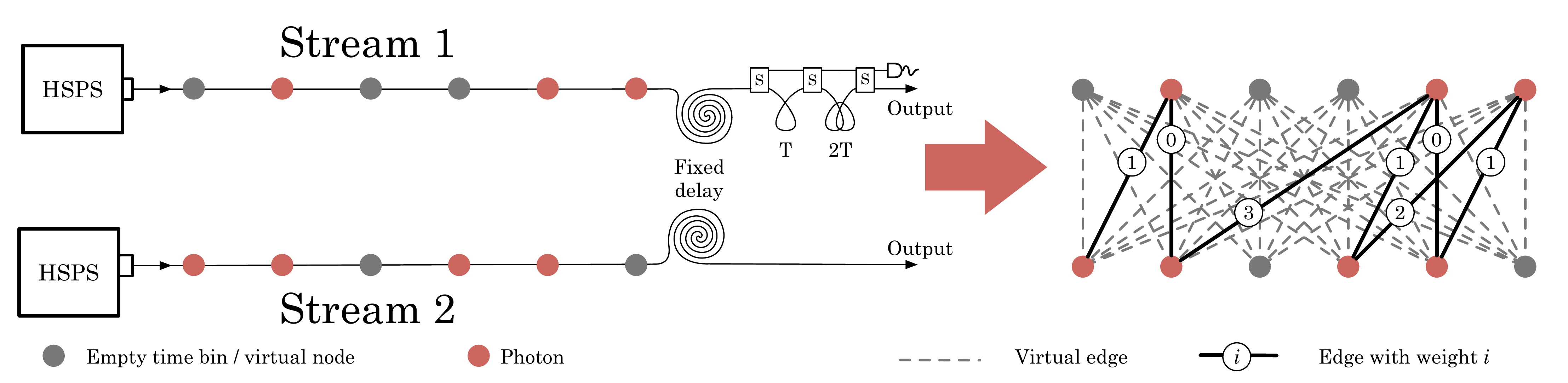}
\caption{{\it The RMUX problem for two streams translated into a the graph assignment problem:}  \textcolor{black}{The photons are generated by HSPSs, and therefore their presence in the stream is known.} Photons on stream 1 can be delayed using a binary-delay network; two switches are required to achieve variables delays between 0 and 3, and an additional switch is required to select the correct output port.  The corresponding bipartite graph has edges with weight corresponding to the required delay for each matching.  Matches which are unachievable are represented by virtual edges with a weight several orders of magnitude larger than any normal edge.}
\label{fig:MatchingProblem}
\end{figure}

The problem here can be rephrased as a bipartite-graph matching problem, where the matching with lowest total edge weight is sought.  A bipartite graph is one in which the vertices can be separated in two disjoint sets, with each edge (which might be weighted) connecting one vertex from each set.  An optimal matching is typically defined as a set of edges without common vertices, with the largest possible number of edges, and with an overall weight that is maximised or minimised (depending on the problem). In the problem of synchronisation of two streams of photons, each stream corresponds to one of the disjoint sets of vertices, and the possibility of synchronising pairs of photons corresponds to an edge. Each edge is assigned a weight which is the number of time bins the photon in the first stream (which is the one with the binary-delay network) has to be delayed in order to be synchronised with the photon in the second stream.

The problem of finding an optimal matching of two photon streams, thus phrased as a bipartite-graph matching problem, can be understood as an instance of the so-called \emph{assignment problem}, can be stated as follows: Let $D = (d_{i,j})$ denote an $n \times n$ matrix of non-negative integers, with $d_{i,j}$ being the weight of the graph edge ${i,j}$ between vertices $i$ and $j$. The optimal solution is found when a set of $n$ independent elements of $D$ is chosen such that no two elements lie in the same row or column of the matrix, and the sum of these elements is minimised.  The assignment problem has been well studied in the literature and a polynomial runtime algorithm was proposed by Kuhn \cite{Kuhn1955} and simplified by Munkres \cite{Munkres1957}; it is often referred to as the Hungarian algorithm. The number of operations required to find the optimal solution scales as $O(n^3)$, where $n$ is the number of vertices in each disjoint set, or equivalently, the number of columns of the matrix.

To make use of the algorithm proposed by Munkres, we choose the entries of the $D$ matrix to correspond to the edge weights for delays in stream 1, and for cases where there is no regular edge, meaning that it's impossible to synchronise the photons, we introduce a virtual edge with a very large weight.  The weight given to the the virtual edges must be several orders of magnitude larger than the largest regular edge weight, to prevent the Munkres algorithm from finding optimal matchings involving virtual edges.  Furthermore, it is often the case that the two streams do not have the same number of photons. To address this, virtual vertices are added to the graph with virtual edges to all vertices for the other stream.  Once the optimal matching has been found, pairings that include virtual nodes or edges are discarded.  An example of the transformation of the problem of synchronising photons in two streams into an assignment problem for a bipartite graph is illustrated in figure \ref{fig:MatchingProblem}.

\begin{figure}[hbt]
\includegraphics[width= 0.7\linewidth]{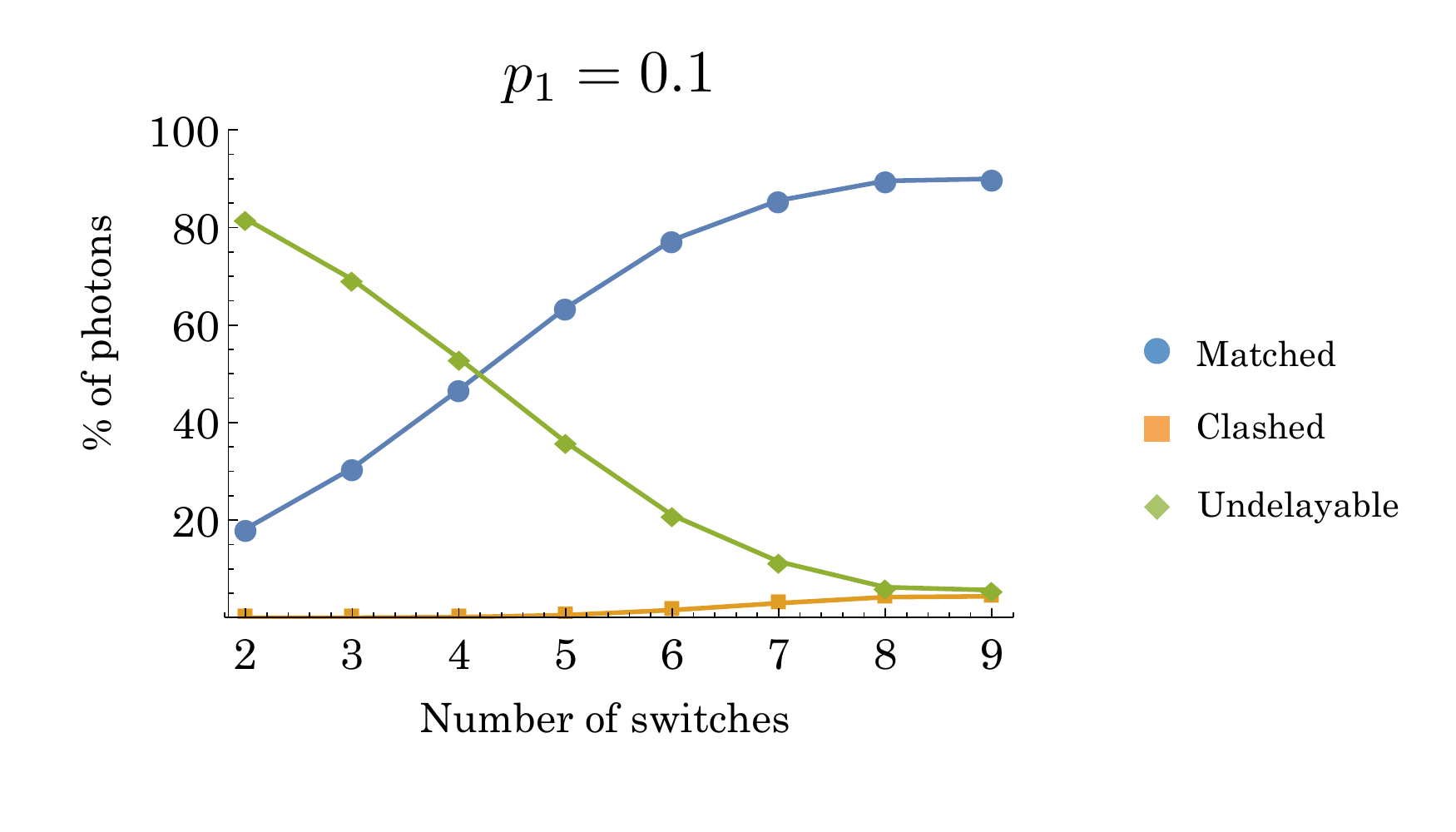}
\caption{{\it Achievable matching using RMUX on two streams:} Photons are generated in each stream with probability $p=0.1$.  The blue line shows the percentage of photons that can be matched on average with another stream as a function of the switch depth for a binary-delay network.  The Hungarian matching method was used to find the matchings, and an optimised algorithm was used to minimise the effects of clashes.  The orange line shows the occurrence of clashes, which is only significant for larger numbers of switches.  The green lines shows the percentage of photons that cannot be matched due to the maximum delay available.\label{fig:matches}}
\end{figure}

At odds with the original assignment problem however, the binary delay network we use for RMUX cannot always achieve the optimal delays for multiple matchings simultaneously, as specified by the output from Munkres's algorithm.  This is because certain configurations of switching delays can be physically incompatible. More specifically, it can be the case that a particular choice of concurrent matchings can require two photons to be incident at both input ports of a switch while demanding opposite switch settings, and we refer to these cases as ``clashes''.  To exclude these clashings, a subroutine is performed after the optimal matching has been found by the Munkres algorithm to identify any clashes. The subroutine finds the subset of graph edges that, once removed, allows for the best matching for the remaining graph.  To quantify the optimal matchings possible in our problem, we have performed Monte Carlo simulations which find the lowest weight matchings possible for random samplings of the streams, while avoiding clashes.  As can be seen from figure \ref{fig:matches}, the rate of clashes for instances with a low number of switches is relatively insignificant.

\begin{figure}[htb]
\includegraphics[width= 0.75\linewidth]{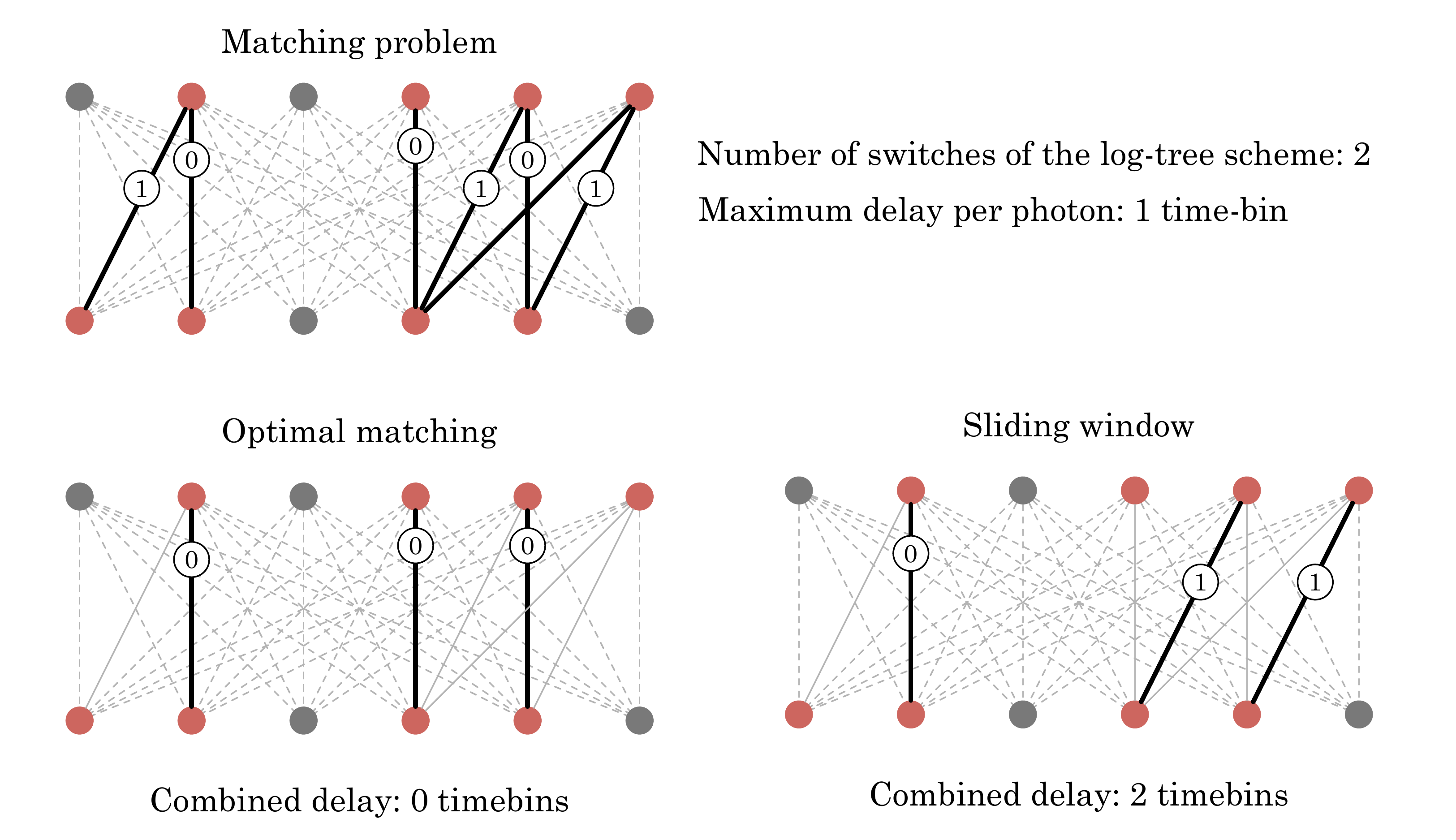}
\caption{{\it Instance of matching solution given by the Hungarian algorithm and by the sliding window strategy:} Note that clashes in switch settings can be avoided for both cases.  The sliding window strategy finds the same number of matches but with an overall higher edge-weight, with the asymmetry of the problem playing in its favour. The top line of vertices of the graph represents the photons that can be delayed, hence there can be vertical edges and edges that connect top-right to bottom-left, but not top-left to bottom-right.
\label{fig:HungarianvsTerry}}
\end{figure}

Unfortunately, implementation of the matching process analysed above presents practical problems: It requires a passive delay for "lookahead" with sufficient length to find optimal matchings, and the classical processing needed to determine these matchings has a substantial overhead that must be dealt in real time using ultra-fast signal processing.  These considerations reveal the need for a simpler matching strategy that requires a short lookahead and simple processing, while achieving near-optimal matching.  We now study one such strategy which shows similarly very good performance, although the average total delay time increases, due to the lack of optimisation over minimum edge weights.  For this strategy, which we call ``sliding window", each photon from stream one is paired with the photon of stream two that is connected to it by the lowest edge weight. Due to the asymmetry of the graph, this photon will be the one produced in the closest time bin. An intuitive way of understanding this strategy, is to think of a window which starts at the first available photon in stream one and catches all time-bins to which that photon can be delayed to. The matching is performed by choosing the photon from stream two which is closest to the photon in stream one at the front of the window. Once this matching is done, we slide the window to the next available photon in stream 1 and repeat the procedure. In figure \ref{fig:HungarianvsTerry}, we can see an instance of the matches produced by both the optimal and the sliding window strategies as well as the total edge weight of the matching (which corresponds to the overall delay).

\begin{figure}[t]
\includegraphics[width= \linewidth]{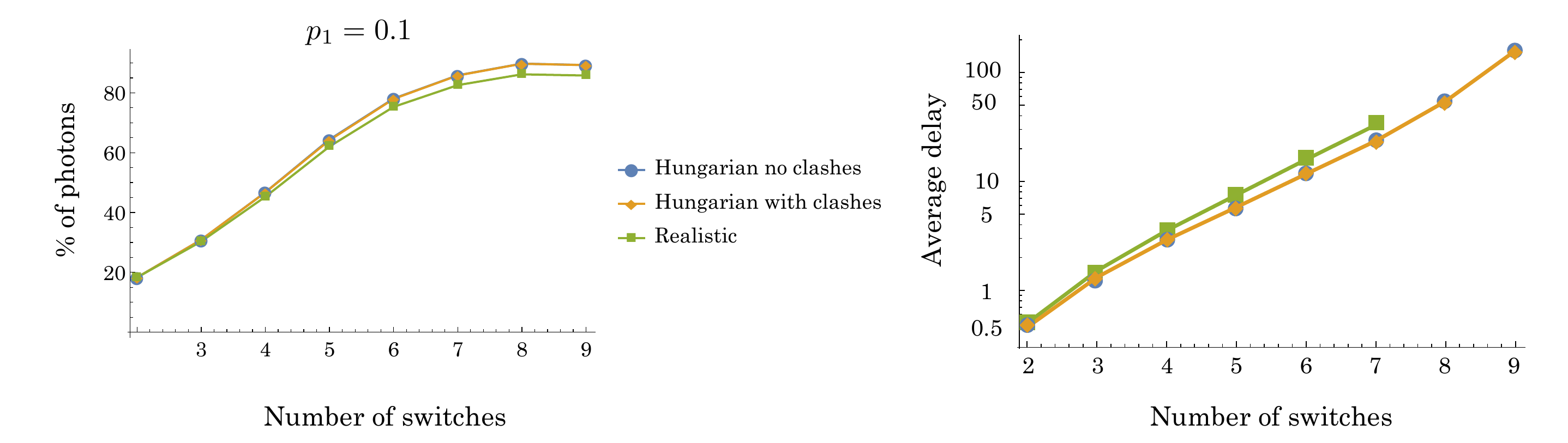}
\caption{{\it Comparison of performance for three matching strategies:} The ``Hungarian no clashes'' strategy shows the best performance achievable using the Hungarian algorithm with no constraints (blue). The ``Hungarian with clashes'' strategy takes into account potential clashes due to incompatible switch settings in the binary-delay network (orange).  Using an effective algorithm for managing the clashes, the effect of clashes can be made negligible. However, our algorithm for this is not suitable for real-time implementation. The third strategy, ``realistic'', uses the ``sliding window" strategy for making the photon pairings, and in the case of any clashes it simply eliminates one of the conflicting matchings. Each data point in the graphs is an average over the output of a Monte Carlo simulation over $100$ sets of stream data (and the three strategies were tested using the same data).\label{fig:comparison}}
\end{figure}

To have a realistic efficient algorithm that can be used in an experiment when using the sliding window strategy, we also use a much simpler clashing algorithm than the one used for the Hungarian algorithm in order to reduce the total length of the delay lines.  Whenever there is a clash, we simply throw away one of the pairs of photons involved in the clash, rather than trying to find a different matching with less clashes. Figure~\ref{fig:comparison} compares the results of three strategies: Hungarian matching ignoring clashes (the provably optimal strategy), Hungarian matching with smart treatment of clashes, and a realistic strategy which is sliding window with inefficient management of clashes. As can be seen from the figure, the realistic strategy does not perform much worse when the switch count is low, which is, in any case, the regime of most importance for experiments.

\begin{figure}[hbt]
\includegraphics[width=\linewidth]{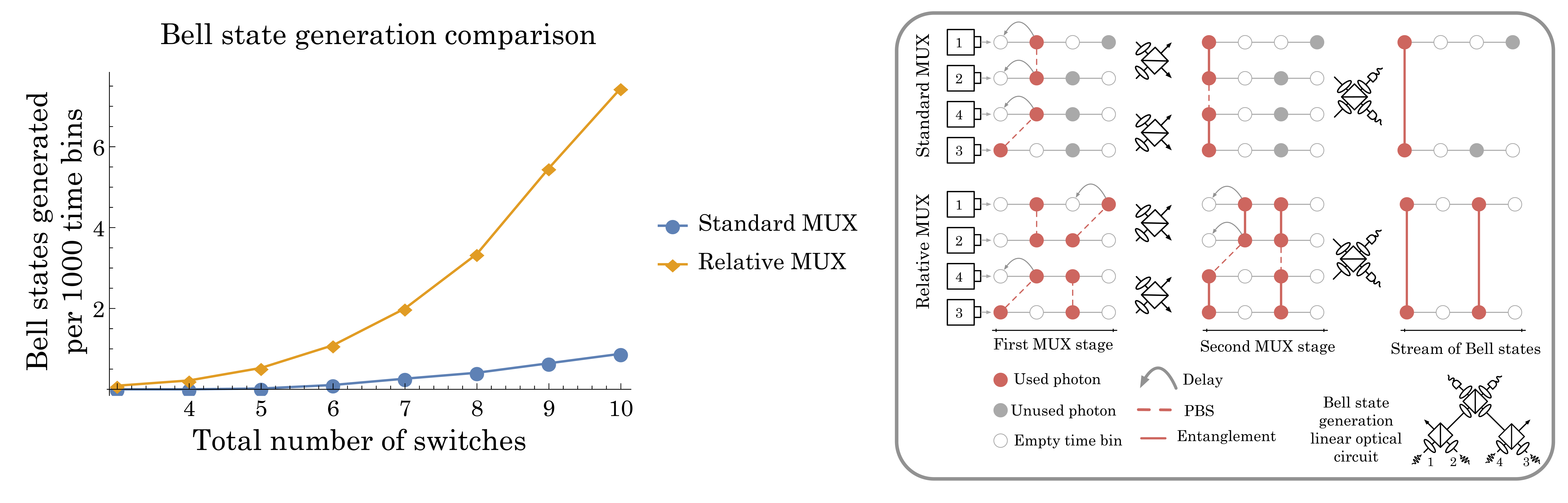}
\caption{{\it Performance comparison of standard MUX and RMUX in the context of Bell State generation:} In the graph on the left, we compare the number of Bell states generated per attempt in a concatenated multiplexing scheme using standard MUX and RMUX, for a source of $p_1=0.1$. Each point in the graphs has been generated by a Monte-Carlo simulation of the multiplexing process, with $10^2$ repetitions. The figure on the right represents a schematic view of each multiplexing process and how it has been simulated. Once all the matches have been considered, only an eighth of the generated matches are kept, to take into account the probabilistic nature of the Bell state generator.  The switch count used represents the total number of switches for both multiplexing stages, the results shown are the optimal for each total number of switches.\label{fig:BScomparison}}
\end{figure}
{\it RMUX and standard multiplexing compared in the context of Bell state generation:} The potential of RMUX can be most easily seen when used in practical scenarios where there are several multiplexing stages, for example, in the case of the generation of a Bell state from four single photons. Adopting similar notation to Section II, we write $p_1$ for probability for generating single photons from heralded source, and $p_2$ for the success probability for generating Bell pairs given four single photons at the input. The scheme we study here is the simpler version proposed in \cite{Zhang2008} (without the additional switch), that generates a Bell state with $p_2=\frac{1}{8}$. In the simulation, we generate four streams of photons with probability $p_1=0.1$. As in the case of 3-GHZ states explained in table \ref{table:GHZMUX}, there are two multiplexing stages, one from HSPS and a second from single photons to Bell states. The number of time bins used for this simulation is varied over a range for both multiplexing stages and the results shown in figure \ref{fig:BScomparison} show the best performing configuration for each total number of switches. In the case of standard MUX, photons are synchronised to the front of the window before passing to the second stage of multiplexing, and only one of the generated photons per stream per window is used in each instance. In contrast, in RMUX all possible pairs of photons are considered and passed to the next multiplexing stage. We also take into account the failure probability of the gate, which has the effect of lowering the success probability at the end of each stage. We can see that for low switch counts, both standard MUX and RMUX produce less than one Bell pair per 1000 time bins, but as the total number of switches increases, RMUX becomes vastly more efficient.


\section{Application of RMUX in a LOQC architecture}
\label{sec:BallisticArchitecture}

{\it Architecture based on percolation of a cluster state on the diamond lattice:}  In this section we explore how RMUX can be applied to an architecture for LOQC to achieve significant reductions in demands for active switching.  We focus on an architecture based on the proposal set out in \cite{Gimeno-Segovia2015}, which builds a 3D cluster state from 3-GHZ resource states.  The idea of the proposal is to create a 3D lattice by fusing the resource states using variants of the type-II fusion gates, which were originally introduced in \cite{Browne2005}.  Although these fusion gates are non-deterministic, a 3D lattice can be generated provided the success probability of the gates exceeds the corresponding percolation threshold.  Reference \cite{Gimeno-Segovia2015} considers the diamond lattice together with so-called ``boosted'' fusion gates.  These gates use ancilla photons to achieve an increased success probability of $75\%$ \cite{Grice2011,Ewert2014}, and thereby operate above the corresponding percolation threshold which is $62.5\%$.  Of particular importance for the following discussion is the fact that this threshold was shown in \cite{Gimeno-Segovia2015} to exhibit a robustness to photon loss.  In section \ref{sec:PercolationResults} we will argue how RMUX can be used to improve the loss threshold further.

\begin{figure}[hbt]
\centering
\includegraphics[width=0.5\linewidth]{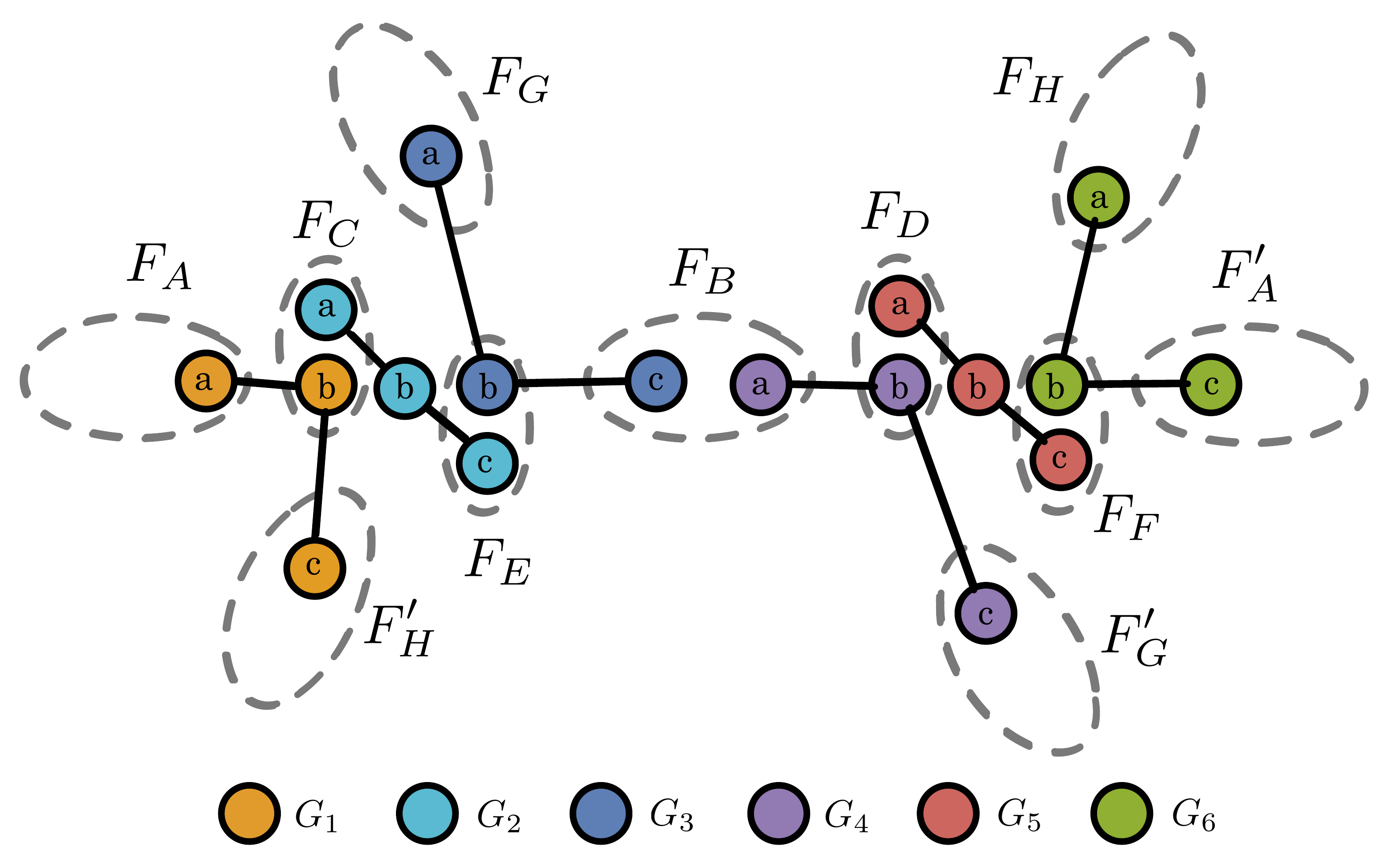}
\caption{{\it Generation of a unit cell within the diamond lattice:} Fusion operations $F_C$ and $F_E$ act on $\{G_1,G_2,G_3\}$ to generate the first microcluster, and operations $F_D$ and $F_F$ act on $\{G_4,G_5,G_6\}$ to generate the second microcluster; $F_B$ fuses the two microclusters to generate a unit-cell cluster.  Connectivity to adjacent unit cells within the same 2d time slice is attempted by fusions $F_A^{(')}$ and $F_G^{(')}$ while fusions $F_H^{(')}$ generate connectivity to the subsequent (previous) time slice.  Note that $'$ denotes operations associated with an adjacent unit cell, and that the colour scheme and labelling is shared with figure~\ref{fig:UnitCellOnChip} in the Appendix. \label{fig:relative fusions}}
\end{figure}

To focus on the part of the LOQC architecture that is relevant to our discussion, we consider the generation of a ``unit cell'' which makes up the diamond lattice, as illustrated in figure~\ref{fig:relative fusions}. \textcolor{black}{ The unit cell is made by fusing together six 3-GHZ states, labelled $\{G_1,\cdots,G_6\}$, which are generated in a heralded non-deterministic way \cite{Varnava2008} and hence require some form of active switching before being directed to the fusion gates.}  In principle, each unit cell can support up to two logical qubits in the final lattice; the fusion operations remove the photons on which they act while (probabilistically) creating connectivity within the lattice.  Following reference \cite{Gimeno-Segovia2015}, we further simplify by assuming that each unit cell is created by fusing together two micro-clusters (five-qubit star clusters), formed in turn from GHZ states $\{G_1,G_2,G_3\}$ and $\{G_4,G_5,G_6\}$.  The operations and delays on the photons in $\{G1,G2,G3\}$ are same as for $\{G_4,G_5,G_6\}$, they only differ in their connections to other microclusters.  By using RMUX, many delay lines that would formerly have been actively switched are now made completely passive, since only one of the photons per fusion operation will be actively switched.  Details of the on-chip arrangement of fusion operations and delay lines (passive and active) required to build the final cluster state using RMUX are given in the Appendix in figure~\ref{fig:UnitCellOnChip}.

\begin{figure}[b]
\begin{center}
\includegraphics[width=0.6\linewidth]{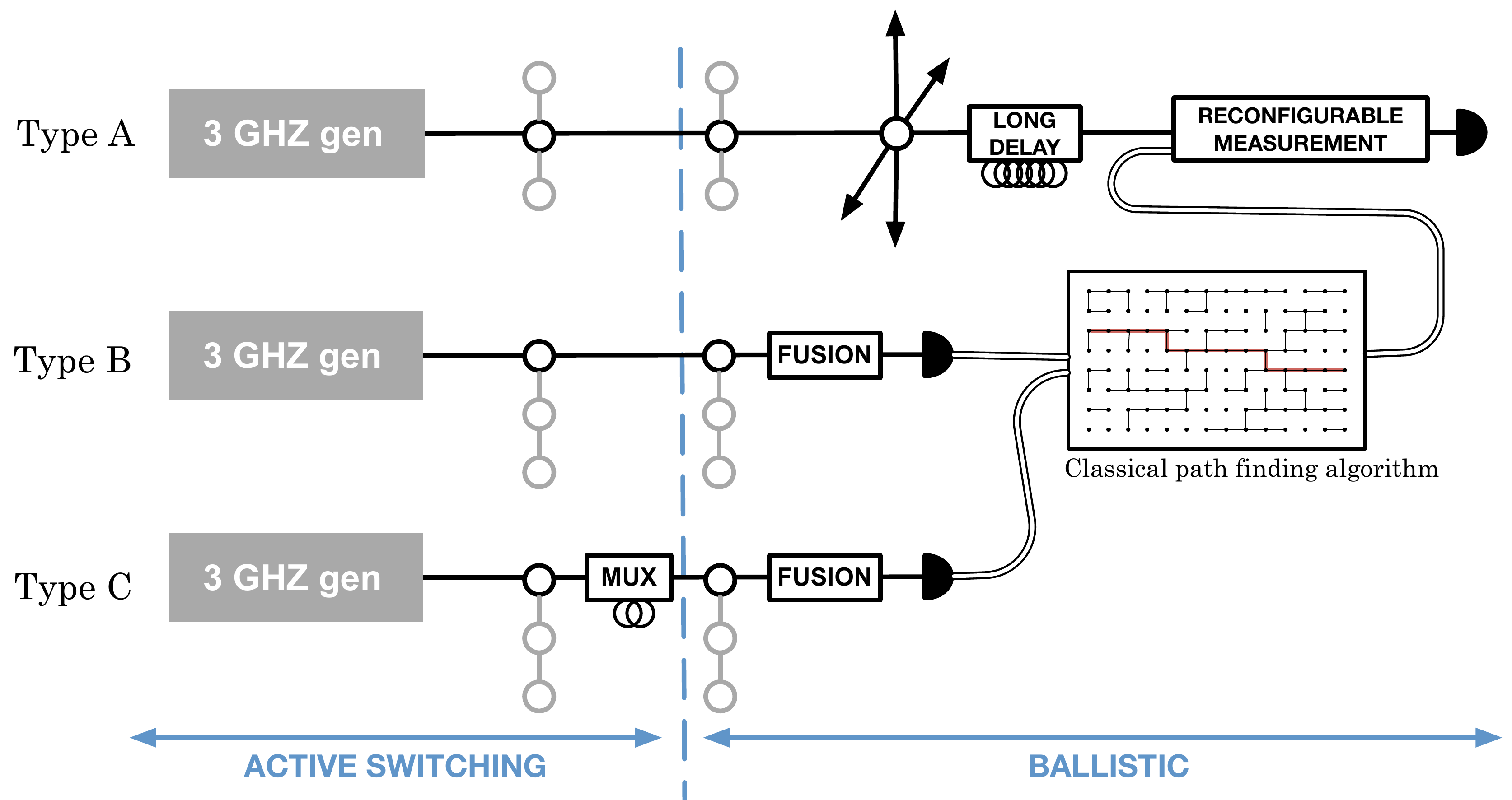}
\caption{{\it World lines of type A, B, and C photons}: The world lines of the photons in GHZ resources states in an implementation of the proposal of \cite{Gimeno-Segovia2015} using an RMUX approach.  This schematic illustrates the operations that each type of photon are subjected to in the architecture. The measurement outcomes of the fusion operations are fed as classical information to a path-finding algorithm, which allows for MBQC on the percolated lattice.  The information is used to choose the basis for a  (reconfigurable) measurement on each data qubit.
\label{fig:worldlines}}
\end{center}
\end{figure}

We can classify the photons from the GHZ states in three classes depending on the operation performed on them, and we will label them Types A, B and C as follows:
\begin{itemize}
\item Type A photons become part of the final cluster, i.e. the data qubits.  To this type belong photons labelled as $G_2(b)$ and $G_5(b)$. These do not need to go through any multiplexing, and only pass through one switch for the final measurement, as part of the MBQC protocol. Note, that this type of photon will have to go through a long passive delay to allow time for the classical processing (for percolation, MBQC and quantum error correction) to determine the right measurement setting.
\item Type B photons will be measured in the fusion operations, but will not be actively delayed. To this class belong photons $G_1(c)$, $G_2(a)$, $G_2(c)$, $G_3(c)$, $G_4(c)$, $G_5(a)$, $G_5(c)$ and $G_6(c)$.
\item Type C photons will be measured in the fusion operations, and must be multiplexed to achieve the correct time bin using RMUX. To this class belong photons $G_1(a)$, $G_1(b)$, $G_3(a)$, $G_3(b)$, $G_4(a)$, $G_4(b)$, $G_6(a)$ and $G_6(b)$.
\end{itemize}
The situation is illustrated in figure~\ref{fig:worldlines}, which shows a schematic for the LOQC architecture with ``world lines'' for the three types of photons, indicating the linear-optical elements encountered between source and detector.  Measurements are  performed on data qubits in the percolated lattice to implement MBQC. \textcolor{black}{It is worth noting that photons of all three types will be subject to a constant loss rate related to the optical operations they have undergone prior to the 3-GHZ generation. In this manuscript, we do not consider this constant rate, as it would depend heavily on the experimental details of the generation of the 3-GHZ state and any number that we could provide would be meaningless without a thorough study of the experimental process that permits the generation of the 3-GHZ state, which is outside the scope of this paper. Our results highlight the improvement in the amount of loss per photon that this LOQC architecture is able to tolerate due to the improved switching scheme, which has a direct relation to the experimental requirements needed for the elements of the switching network.}

\section{ Loss tolerance results for percolation}
\label{sec:PercolationResults}

Next we revisit the loss-tolerance results reported in \cite{Gimeno-Segovia2015}, which apply to the mixed (i.e. site and bond) percolation threshold for the diamond lattice with unit cell as in figure~\ref{fig:relative fusions}.  The original results of \cite{Gimeno-Segovia2015} assumed that all photons from the 3-GHZ states which are involved in fusion operations are subject to losses, as are the ancilla photons used to boost the success probability for the fusion operations.  The same loss rate is assumed for all photons from GHZ states as well as the ancillae photons.  These assumptions could apply when standard MUX is used to achieve sychronisation of all photons which are input into fusion measurements, where all loss arises from switches in the active delay networks (this reflects the $~10\times$ difference in typical loss rates for active and passive components used in photonic circuitry \cite{Rambo2013,Zhang2013,Sheng2012}).  Note that for simplicity the 3-GHZ generators themselves are assumed to have deterministic (lossless) single-photon inputs \cite{Eisaman2011}, and therefore generate 3-GHZ states with probability $1/32$ \cite{Varnava2008}.

Here we consider how the threshold for percolation compares for a loss model based on the RMUX-implementation described in section~\ref{sec:BallisticArchitecture} and summarised in figure~\ref{fig:worldlines}.  Following figure~\ref{fig:worldlines}, we now distinguish between Type A and B photons where we disregard losses, Type C photons which are actively delayed using RMUX before fusion operations and which are subject to loss with rate $p_l$, and ancilla photons that are subject to loss with rate $a_l$ (which arise due to the need for active synchronisation with the GHZ states at the input of the fusion operations).  For the fusion gate, we take the scheme in \cite{Grice2011} which uses a Bell pair for the ancilla (this uses half the number of ancilla photons compared to the scheme in \cite{Ewert2014} and therefore is the least susceptible to loss).
To account for this loss, we adopt a simplified model: if the number of photons detected at a boosted fusion gate is less than the number expected, the fusion is counted as failed on account of loss \footnote{This is the simplest model, as it does not consider differences in loss tolerance between the two boosted fusion gates or events that can be considered successful despite the loss of a photon.}.  The loss tolerance for percolation will then depend on the probability for a fusion operation suffering a loss, $f_l$, which is given by $f_l=1-(1-p_l)(1-a_l)^2$; when there is no loss, the success probability for fusion is $75\%$.

\begin{figure}[t]
\begin{center}
\includegraphics[width=0.5\linewidth]{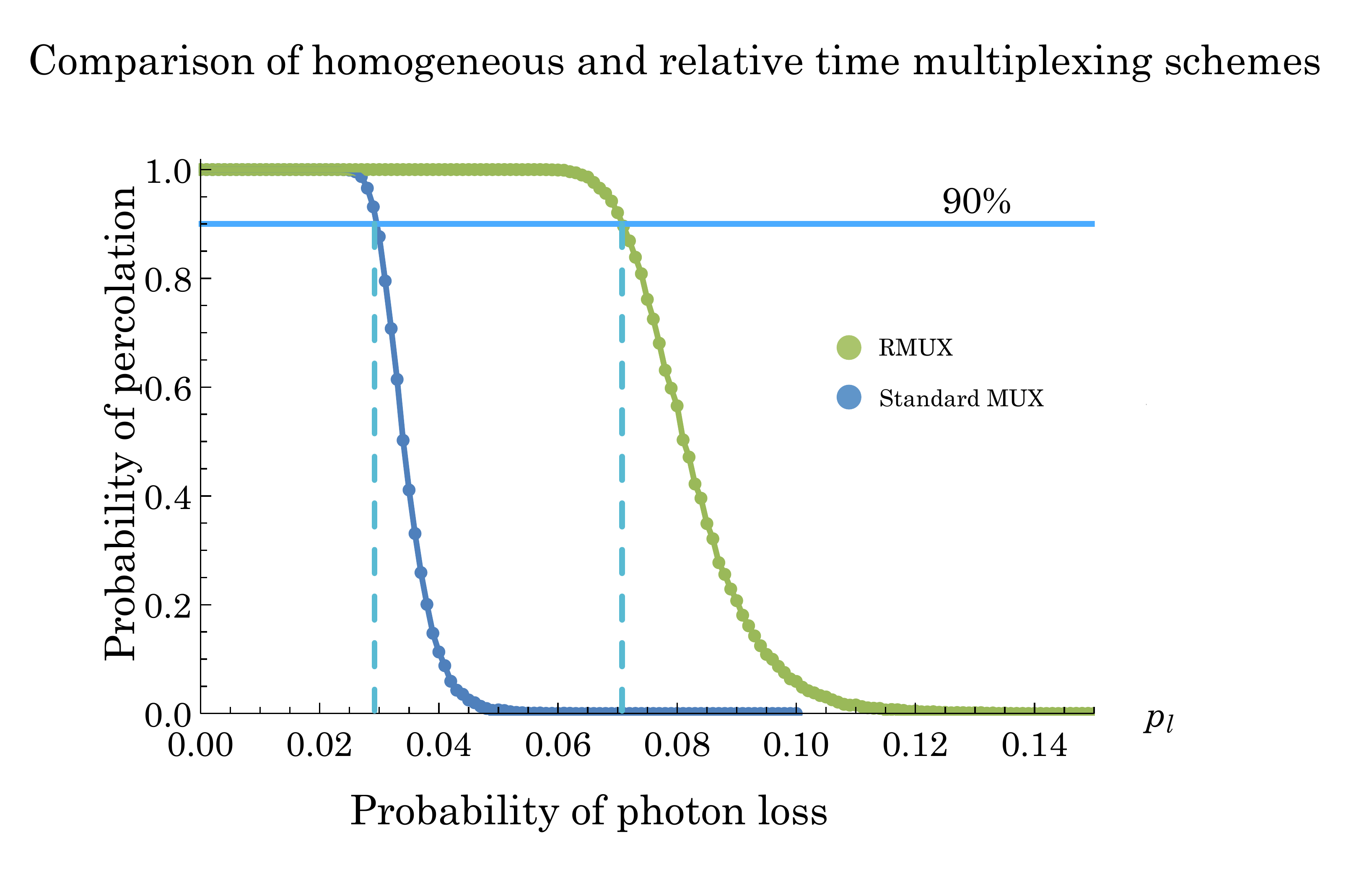}
\caption{{\it Comparison of loss tolerance for percolation with no loss on ancilla photons:} Comparison of tolerable loss for standard MUX and RMUX implementations of \cite{Gimeno-Segovia2015}. The use of RMUX scheme boosts the tolerable loss to more than twice what can be tolerated with the standard MUX scheme.
\label{fig:noancilla}}
\end{center}
\end{figure}

Our ability to perform universal quantum computation depends on the percolation properties of the lattice as explained in \cite{Gimeno-Segovia2015}.
Mixed percolation thresholds can be obtained numerically to incorporate the effect of photon loss on the percolation properties of the lattice.  The percolation simulations we perform account for different types of outcomes for the fusion operations: success outcomes, failure outcomes without loss which can still create lattice connectivity, and failure outcomes due to photon losses.  We first assume that ancilla photons are lossless, and compare both standard MUX and RMUX implementations. Numerical results are shown in figure \ref{fig:noancilla}. It can be observed that for a percolation probability $\ge 90\%$, the RMUX implementation can tolerate up to $7\%$ photon loss, while the standard MUX implementation with equal losses on all GHZ qubits undergoing fusion can only tolerate $2.9\%$. Note that this last result is compatible with the $1.6\%$ tolerable loss rate reported in \cite{Gimeno-Segovia2015}, where the loss rate of all GHZ and ancilla photons was assumed equal, while figure \ref{fig:noancilla} assumes lossless ancilla photons.

\begin{figure}[t]
\begin{center}
\includegraphics[width=0.8\linewidth]{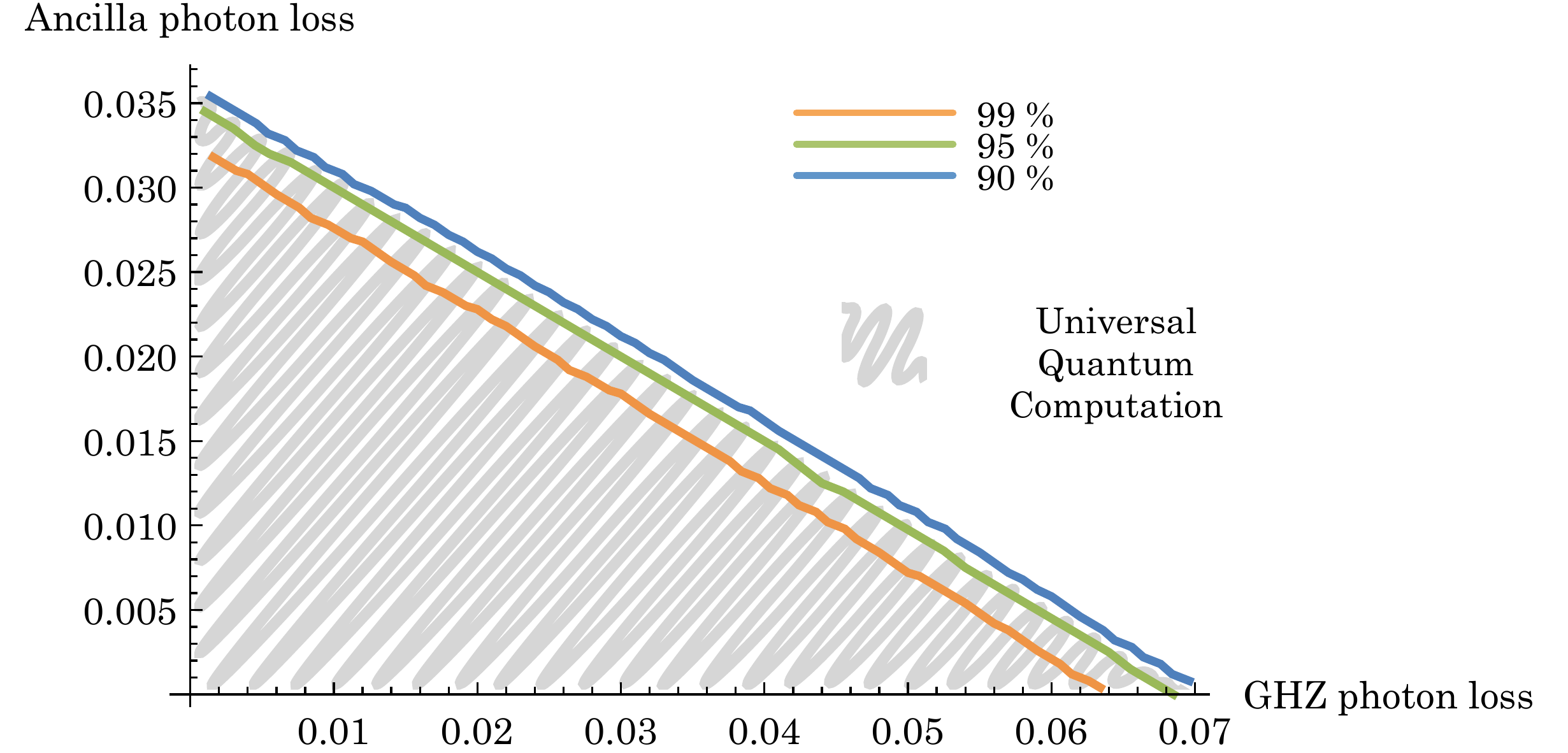}
\caption{{\it Loss threshold trade-off with ancilla-photon and GHZ-photon loss:} The grey shaded area highlights the area of phase space where universal quantum computation is possible using the RMUX implementation of \cite{Gimeno-Segovia2015}. We have marked the threshold for different percolation probabilities.
\label{fig:lossthresholdtradeoff}
}
\end{center}
\end{figure}

However, it is not realistic to assume that the ancilla photons used to boost the fusion are lossless.  In figure~\ref{fig:lossthresholdtradeoff}, shaded in grey, we can see the range of values of photon loss and ancilla loss that can allow us to perform universal quantum computation using the RMUX implementation. We have marked three different thresholds depending on what percolation probability is desired, $90\%,\,95\%$ or $99\%$. It might be surprising that the threshold is linear, given the non-linear dependence of fusion loss rate $f_l$ on $p_l$ and $a_l$. $f_l$ is indeed not linear with respect to the individual loss rates, but for the range of photon loss of interest, the leading term in the expansion of $f_l$ is a linear term dependent on $p_l+2a_l$, and the rest of the terms are negligible (accounting for a maximum of $5\%$ of $f_l$).

\section{Conclusions and Outlook}
\label{sec:Conclusions}

In this paper, we have provided evidence that techniques under the umbrella of RMUX can be used to achieve large savings in complicated
linear-optic circuits.  RMUX can be harnessed to exploit the fact that determinism in LOQC architectures does not require all (or even most) events to be synchronised.  The savings achievable using RMUX are particularly important in light of the technological difficulties for achieving fast, low-loss and high-efficiency switching.  By facilitating improvements in resource consumption and efficiency, RMUX-based implementations of LOQC architecture can allow large relaxation of performance demands at the component level.  There are very immediate extensions of the calculations discussed in this paper.  For example it is straightforward to include the effect of additional losses from passive elements and delays, as well as to compare performance using different types of (boosted) fusion gates and alternative schemes for generating Bell or GHZ states.  The analysis can then be extended to compute allowable tolerances for switch components based on alternative designs for the multiplexing networks that would be required.

Although our discussion has been limited to examples involving the generation of entangled states and percolation on a diamond lattice, it is clear that a similar approach can be applied at all levels of any LOQC architecture.  One key realisation is that, if we could use the resources to their full potential, this would have a big impact on the LOQC strategy proposed in \cite{Gimeno-Segovia2015}.  The examples we have given in this paper represent conservative applications of RMUX.  Future work will explore if RMUX can be used to enable near-deterministic generation of full micro-clusters: in this scenario, the percolation properties of the diamond lattice would only depend on the ability to create bonds between the micro-clusters, representing a move to a pure bond percolation rather than a mixed site-bond percolation.  More speculatively, it would be enlightening to explore schemes where whole regions of a percolating lattice are grown asynchronously.

\section{Acknowledgements}
This work was supported by the UK Engineering and Physical Sciences Research Council (EPSRC).

\appendix

\section{Unit-cell generation on chip}
\label{sec:Appendix}

\begin{figure}[h]
\begin{center}
\includegraphics[width=0.92\linewidth]{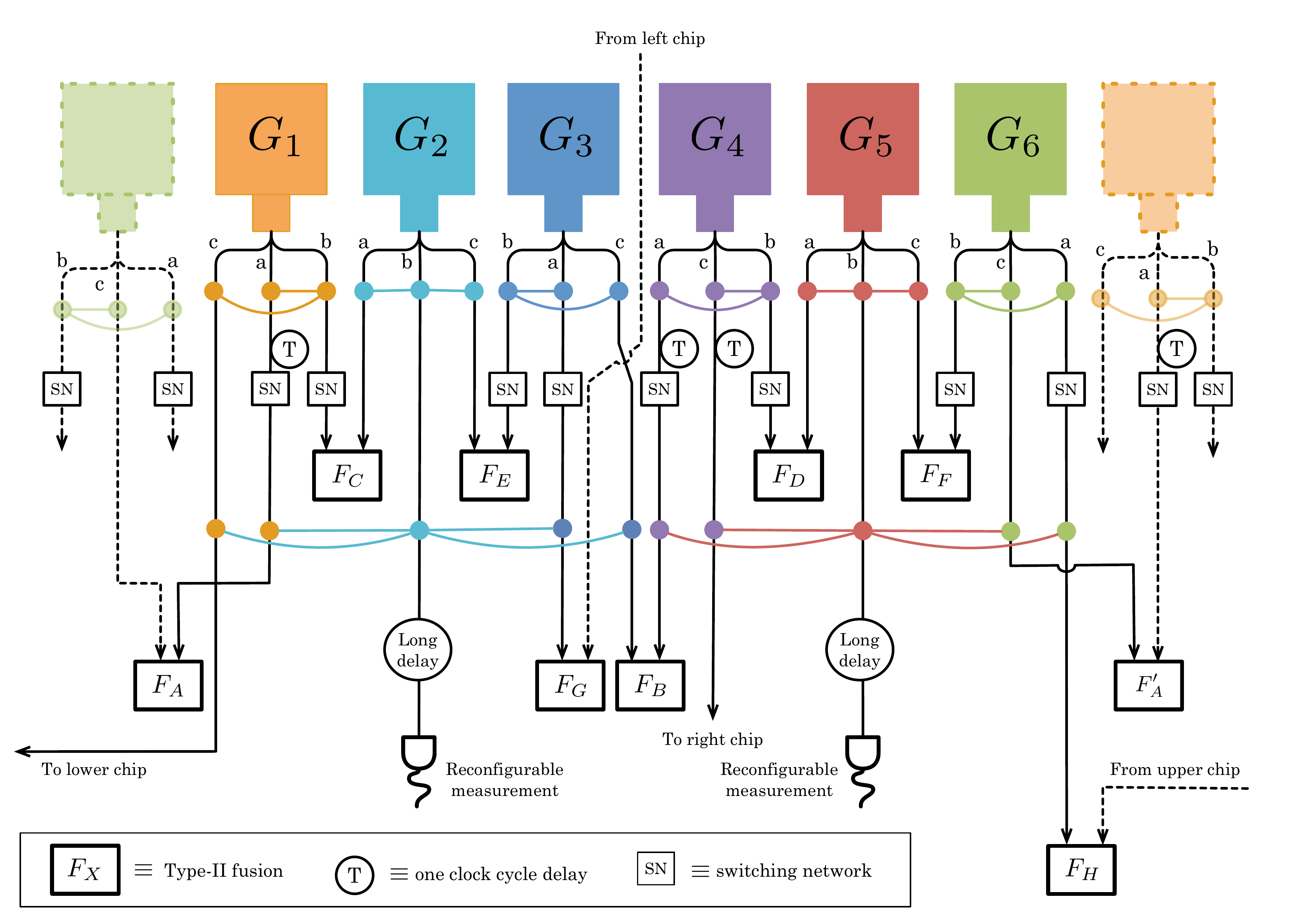}
\caption{{\it Arrangement of 3-GHZ generators and fusion operations on chip to generate a single unit cell within a diamond lattice:} The notation of this figure follows that of figure~\ref{fig:relative fusions}. The layout presented here would be repeated in a wafer, with photonic chips interconnecting as shown. Short delays of one clock cycle to synchronise fusions are marked with the letter T, while the switching networks required for the RMUX are marked with SN.  The dashed lines represent elements from nearby unit-cell generators, while thick lines mark the photons that benefit from the RMUX scheme by not passing through any switches.  The photons which undergo the long delays are the \emph{data} qubits. Each data qubit is subject to only one active element for the final measurement (which must be reconfigurable to implement standard the MBQC protocol).
\label{fig:UnitCellOnChip}}
\end{center}
\end{figure}


\providecommand{\newblock}{}

\end{document}